\documentclass[prb,showpacs,amsfonts,amsmath,floatfix,twocolumn]{revtex4} 

\usepackage{graphicx}
\usepackage{epstopdf}
\usepackage{amssymb}
\usepackage{amsmath}
\usepackage{xspace}
\usepackage{color}
\usepackage{xcolor}

\newcommand{\mean}[1]{\left\langle #1 \right\rangle}
\newcommand{\bra}[1]{\left\langle #1 \right\vert}
\newcommand{\ket}[1]{\left\vert #1 \right\rangle}

\usepackage{color}
\usepackage{soul}

\begin{document}

\title{Stimulated Raman Adiabatic control of a nuclear spin in diamond}

\author{Raul Coto$^{1}$}
\email{rcoto@uc.cl}

\author{Vincent Jacques$^{2}$} 

\author{Gabriel H\'etet$^{3,4}$}

\author{Jer\'onimo R. Maze$^{1}$}
\email{jmaze@uc.cl}

\affiliation{$^1$Instituto de F\'{\i}sica, Pontificia Universidad Cat\'{o}lica de Chile,
Casilla 306, Santiago, Chile\\$^2$Laboratoire Charles Coulomb, Universit\'e de Montpellier and CNRS, 34095 Montpellier, France\\$^3$ Laboratoire Aim\'e Cotton, CNRS, Universit\'e Paris-Sud, ENS Cachan, Universit\'e Paris-Saclay
91405 Orsay Cedex, France\\$^4$Laboratoire Pierre Aigrain, CNRS, Universit\'e Pierre et Marie Curie, Universit\'e Paris Diderot and Ecole Normale Sup\'erieure, 75005 Paris, France}

\begin{abstract}

Coherent manipulation of nuclear spins is a highly desirable tool for both quantum metrology and quantum computation. However, most of the current techniques to control nuclear spins lack of being fast impairing their robustness against decoherence. Here, based on Stimulated Raman Adiabatic Passage, and its modification including shortcuts to adiabaticity, we present a fast protocol for the coherent manipulation of nuclear spins. Moreover, we show how to initialise a nuclear spin starting from a thermal state, and how to implement Raman control for performing Ramsey spectroscopy to measure the dynamical and geometric phases acquired by nuclear spins.

\end{abstract}

\maketitle

\section{Introduction}

Nuclear spins in solid-state systems are leading candidates for long-lived quantum memories and high fidelity quantum operations as they are isolated from the environment due to their relatively small magnetic moment compared to that of electrons. However, in order to enable such applications, this advantage possesses a challenge for accessing and coherently manipulating nuclear spins. Multiple examples of coherent control of nuclear spins have been presented by using hyperfine interactions with an available electronic spin. Ensemble of nuclear spins were accessed using phosphorus electronic spins in silicon \cite{Steger:2012}, whereas individual nuclear spins have been accessed in diamond \cite{Gurudev, Neumann:2010},  silicon \cite{Jarryd:2013}, through an optically accessible ancillary electronic spin, and in single molecules using external electric field control\cite{Thiele:2014}. Several nuclear spins in diamond have been controlled using hyperfine \cite{Taminiau,Waldherr} and nuclei dipole-dipole interactions \cite{Jiang:2009}. Such control has enabled the production of GHZ states \cite{Jiang:2009} and the implementation of error correction in multi-qubit spin registers \cite{Taminiau,Waldherr}. Recently, new methods for controlling a nuclear spin have been proposed by synchronously driving an electronic spin with the nuclear Larmor precession \cite{Mkhitaryan:2015}. Here, we present a method for preparing and controlling a nuclear spin in diamond using Stimulated Raman Adiabatic Passage (STIRAP) in the microwave domain following a recent experimental realisation of Coherent Population Trapping (CPT) \cite{Jamonneau}.

STIRAP can coherently transfer population by adiabatically changing a dark state  \cite{Gaubatz,Bergmann1,Bergmann2}. The same process is also possible through a bright state (b-STIRAP) \cite{Klein, Grigoryan} and both methods have been realized in doped solids \cite{Klein,Goto,Klein2,Ohlsson}, cold atoms \cite{Du}, and quantum dots \cite{Hohenester}, to name a few. It has been implemented for coherent manipulation of states \cite{Du} in logic operations \cite{Beil, Remacle}, quantumness witness detection \cite{Huang} and entanglement generation \cite{Chen}. In quantum metrology, it has been proposed to improve the detection of electric dipole moments using ThO molecules \cite{Panda:2016}, and for mapping light states in to nuclear spins states in optical cavities \cite{Schwager:2010}. In diamond, it has been implemented in the optical domain to control the Nitrogen-Vacancy (NV) centre electronic spin \cite{Golter:2014}, and its geometrical phase \cite{Yale:2016}.

In what follows we propose to use stimulated Raman adiabatic passage to control a nuclear spin in diamond that is strongly coupled to the electronic spin associated to an individual NV colour centre. In Section II we introduce to the system and lambda configuration on which STIRAP is implemented. In Section III we use STIRAP to rapidly and coherently manipulate a nuclear spin and initialise it from a thermally mixed state. Finally, we discuss how to use Raman pulses in order to perform spectroscopy on a nuclear spin and measure its geometrical phase.  

\section{The Model}
\subsection{Nuclear spin based $\Lambda$-scheme}
\begin{figure}[ht]
\centering 
\includegraphics[scale=0.6]{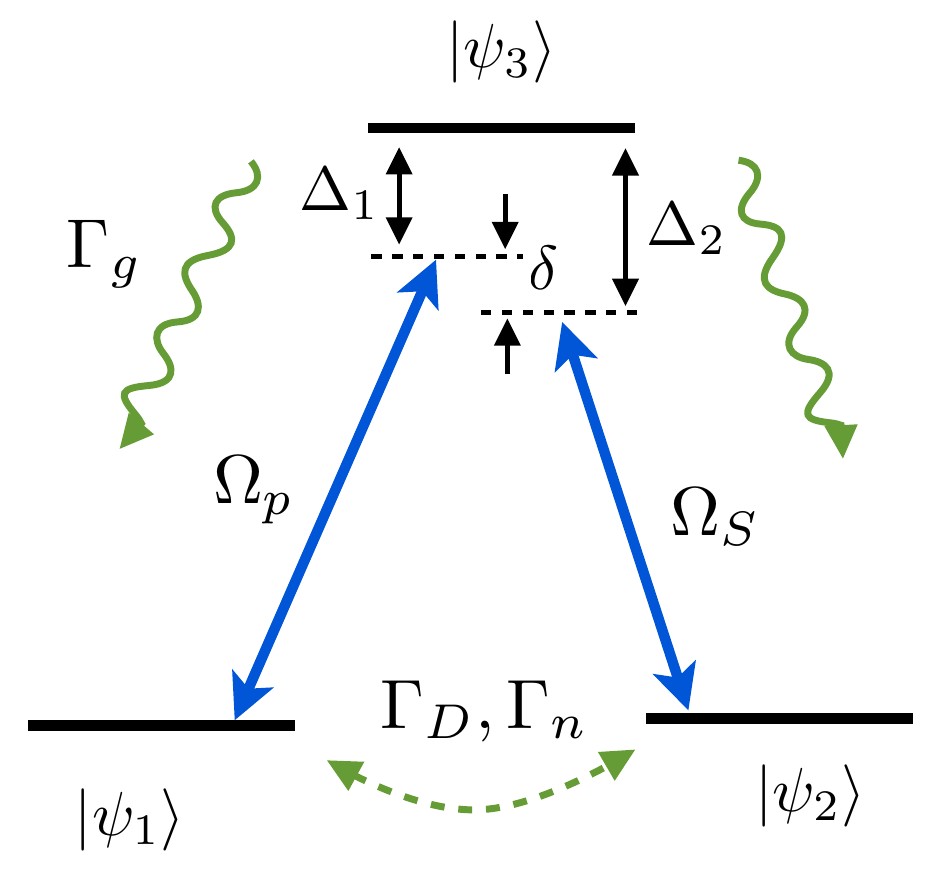}
\caption{(Color online) $\Lambda$-configuration of the three level system. The transitions indicated by the curved and dashed arrows can be highly increased by means of an optical external excitation.}
\label{fig1}
\end{figure}

We consider a Carbon-$13$ nuclear spin in diamond coupled via hyperfine interaction to a nearby NV colour centre which is composed of a vacancy and a Nitrogen substitutional atom (isotope 14). The Hamiltonian describing this nuclear spin and the NV electronic spin is given by ($\hbar =1$)\cite{Jamonneau}

\begin{eqnarray}\label{H0}
H_0 &=& D S_z^2 + \gamma_e S_z B_z + \gamma_n I_z B_z + S_z A_{zz}I_z \nonumber \\&&+ \frac{A_{ani}}{2}S_z(I_+ e^{-i\varphi} + I_- e^{i\varphi}), 
\end{eqnarray}
where $D/2\pi\approx 2.87$ GHz is the zero-field splitting, $\gamma_e/2\pi\approx 2.8$ MHz/G  and $\gamma_n/2\pi\approx 1.07$ kHz/G are the electronic spin and $^{13}C$ nuclear spin gyromagnetic ratio, respectively. The last two terms correspond to the Hyperfine interaction after applying a secular approximation justified by the large value of $D$ compared to the hyperfine tensor components $A_{i,j}$, i.e., we have neglected terms proportional to $S_x$ and $S_y$. $A_{ani}=(A_{zx}^2 + A_{zy}^2)^{1/2}$ and $I_{\pm}=I_x \pm i I_y$. In equation (\ref{H0}) we have assumed a fixed $^{14}N$ nuclear spin projection, e.g., $m_{I_N}=+1$ \cite{Jamonneau}. The eigenenergies of $H_0$ are \cite{Jamonneau,Dreau}: $E_{1,2} = \pm\gamma_n B_z/2 ;  E_{3,4} = D-\gamma_e B_z \mp \frac{1}{2}\sqrt{A_{ani}^2 + (A_{zz}-\gamma_n B_z)^2} $, and the corresponding eigenstates read
 
\begin{eqnarray}
\ket{\psi_1}&=& \ket{\uparrow,0}, \nonumber\\
\ket{\psi_2}&=& \ket{\downarrow,0}, \\
\ket{\psi_3}&=& \cos(\vartheta/2)\ket{\uparrow,-1} + \sin(\vartheta/2)e^{i\varphi}\ket{\downarrow,-1}, \nonumber\\
\ket{\psi_4}&=& -\sin(\vartheta/2)e^{-i\varphi}\ket{\uparrow,-1} + \cos(\vartheta/2)\ket{\downarrow,-1}.\nonumber
\end{eqnarray}
where $\ket{\uparrow}$ ($\ket{\downarrow}$) is the Carbon-13 nuclear spin state $m_I=+1/2$ ($m_I=-1/2$), $\ket{0}$ ($\ket{-1}$) is the electronic spin state $m_s=0$ ($m_s=-1$), and

\begin{eqnarray}\label{vartheta}
\tan\vartheta =\frac{A_{ani}}{A_{zz}-\gamma_n B_z}, \qquad
\tan \varphi = A_{zy}/A_{zx}.
\end{eqnarray}
The magnetic field can be chosen so that the $\Lambda$-scheme has a balanced transition intensity $\vartheta\approx \pi/2$ \cite{Jamonneau}, i.e., $A_{zz} = \gamma_nB_z$. 
We will focus on the submanifold $\{\psi_1,\psi_2,\psi_3\}$ and neglect state $\psi_4$ as we will only consider resonant and red detuned excitations between the ground states $\{\psi_1,\psi_2\}$ and state $\psi_3$. As a result, the system can be described by a $\Lambda$-configuration (see Fig. \ref{fig1}) with Hamiltonian

\begin{equation}\label{H02}
H_0 = E_1\ket{\psi_1}\bra{\psi_1} + E_2\ket{\psi_2}\bra{\psi_2} +E_3\ket{\psi_3}\bra{\psi_3}.
\end{equation}

\subsection{Stimulated Raman Adiabatic Passage}\label{STIRAP_sec}

Stimulated Raman Adiabatic Passage (STIRAP)\cite{Bergmann1,Bergmann2} can successfully transfer the population from one state ($\ket{\psi_1}$) to another ($\ket{\psi_2}$) by an intermediate state ($\ket{\psi_3}$) by driving the transitions $\psi_1 \leftrightarrow \psi_3$ and $\psi_2 \leftrightarrow \psi_3$ (see Fig. \ref{fig1}). The interaction Hamiltonian takes the form ($\hbar =1$)

\begin{equation}\label{H_i}
H_i=\frac 1 2 \left(\Omega_p(t)\sigma_{31}e^{-i\omega_p t} +\Omega_S(t)\sigma_{32}e^{-i\omega_S t}+ h.c.\right),
\end{equation}
where $\sigma_{ik}=\ket{\psi_i}\bra{\psi_k}$, $\Omega_p(t)$ and $\Omega_S(t)$ are Gaussian time-dependent coupling strengths (see Fig. 2 inset) for the pump and Stokes fields, respectively, given by

\begin{eqnarray}\label{Rabi}
\Omega_p(t)&=& \Omega_{13}\cos^2(\vartheta/2) e^{-\frac{(t-t_d/2)^2}{2 \sigma^2}} \\
\Omega_S(t) &=& \Omega_{23}\sin^2(\vartheta/2) e^{-\frac{(t+t_d/2)^2}{2 \sigma^2}},
\end{eqnarray}
with Rabi frequencies $\Omega_{13}=\gamma_{e}\mathcal{B}_p$ and $\Omega_{23}=\gamma_{e}\mathcal{B}_s$, where $\mathcal{B}_{p,s}$ is the amplitude of the driving field. The factor $\cos^2(\vartheta/2)$ ($\sin^2(\vartheta/2)$) gives the intensity of the transition $\psi_1 \leftrightarrow \psi_3$ ($\psi_2 \leftrightarrow \psi_3$), which is inherent to our $\Lambda$-configuration. We set $\Omega_{13}=\Omega_{23}=\Omega_0$, with $\Omega_0/2\pi = 1$ MHz. The time delay between the pulses is $t_d$ and the overlapping time is defined as $\Delta t \approx 6\sigma -t_d$, where $6\sigma$ includes $99.74\%$ of each Gaussian pulse. To achieve maximum fidelity the time delay is optimised to $t_d = \sqrt{2}\sigma$. In the rotating frame the total Hamiltonian is

\begin{equation}\label{H_STIRAP}
\tilde{H}=\delta\sigma_{22}+\Delta_1\sigma_{33} + \frac 1 2(\Omega_p(t)\sigma_{31} + \Omega_S(t)\sigma_{32} + h.c.) ,
\end{equation}
where $\Delta_1= E_3-E_1 - \omega_p$ and $\Delta_2=E_3-E_2-\omega_S$ are the one photon detunings. We set the two-photon detuning $\delta = \Delta_1-\Delta_2$ to zero ($\Delta_1=\Delta_2=\Delta$) unless otherwise specified. The eigenstates of $\tilde{H}$ are \cite{Fleischhauer,Bergmann1}

\begin{eqnarray}
\ket{b_+} &=&\sin\theta\sin\phi\ket{\psi_1} + \cos\phi\ket{\psi_3} + \cos\theta\sin\phi\ket{\psi_2}, \nonumber \\
\ket{d} &=& \cos\theta\ket{\psi_1} - \sin\theta\ket{\psi_2}, \\
\ket{b_-} &=& \sin\theta\cos\phi\ket{\psi_1} - \sin\phi\ket{\psi_3} + \cos\theta\cos\phi\ket{\psi_2}\nonumber \label{b_menos},
\end{eqnarray}
where
\begin{equation}\label{phi}
\tan 2\phi = \frac{\sqrt{\Omega_p^2(t)+ \Omega_S^2(t)}}{\Delta}, \qquad \tan\theta=\frac{\Omega_p(t)}{\Omega_S(t)}.
\end{equation}

The corresponding eigenvalues are $E_{b\pm}=\Delta/2 \pm \{\Delta^2 +\Omega_p^2(t)+\Omega_S^2(t)\}^{1/2}/2 $ and $E_d=0$. Note that the bright eigenstates $\{\ket{b_{\pm}}\}$ are represented by a linear combination of all bare states, while the dark eigenstate $\ket{d}$ has only the contribution of the two lower states. A coherent population transfer between states $\ket{\psi_1}$ and $\ket{\psi_2}$ can take place by varying the Rabi frequencies that effectively change the angle $\theta$. This transfer does not pass through state $\ket{\psi_3}$ provided the evolution is adiabatic\cite{Kuklinski,Gaubatz,Bergmann1}, which requires that the mixing angle $\theta$ varies much slower than the energy difference between eigenstates, i.e.,

\begin{equation}
\dot{\theta}\ll \vert E_{b\pm} -E_d \vert \equiv\Omega_{eff}.
\end{equation}

When the pulses have a smooth shape, an approximate global criterion for adiabaticity can be found: $\Omega_{eff}\Delta t> 10$ \cite{Gaubatz,Bergmann1}. 

\section{Results}

\subsection{Initializing a nuclear spin}\label{polarizing_sec}

We now consider the preparation or initialization of a nearby Carbon-13 nuclear spin. In order to achieve this goal we apply a pump pulse $\Omega_p(t)$ under green laser excitation to prepare the state $\ket{\psi_2}$ from a thermal state $\rho(0)=1/2\left( \ket{\psi_1}\bra{\psi_1} +\ket{\psi_2}\bra{\psi_2}\right)$. When the pump field is switched on, the population rapidly moves from $\ket{\psi_1}$ to the excited state $\ket{\psi_3}$. At the same time, we induce a strong decay at a rate of $\Gamma_g=5$ MHz from $\ket{\psi_3}$ to $\ket{\psi_1}$ and $\ket{\psi_2}$ by applying a time-pulsed external green laser\cite{Jamonneau}, see Fig. \ref{fig1} for $\Omega_S=0$. This optically induced decay rate is a particular feature of our system since it plays the role of the spontaneous emission, preparing the system in a dark state. Nevertheless, one flaw of this approach is that it optically induces both longitudinal $\Gamma_{n,opt}$ and transverse $\Gamma_{D,opt}$ relaxation of the nuclear spin state which might vary for different $^{13}C$ nuclear spins, limiting the final fidelity. We assume these rates to be of the order of $\Gamma_{n,opt} = 150$ kHz and $\Gamma_{D,opt} = 1$ MHz \cite{thesis:Jamonneau}. The evolution passes the population from $\ket{\psi_1}$ to $\ket{\psi_2}$ as shown in Fig. \ref{fig3_polarization}. 

\begin{figure}[ht]
\centering 
\includegraphics[scale=0.4]{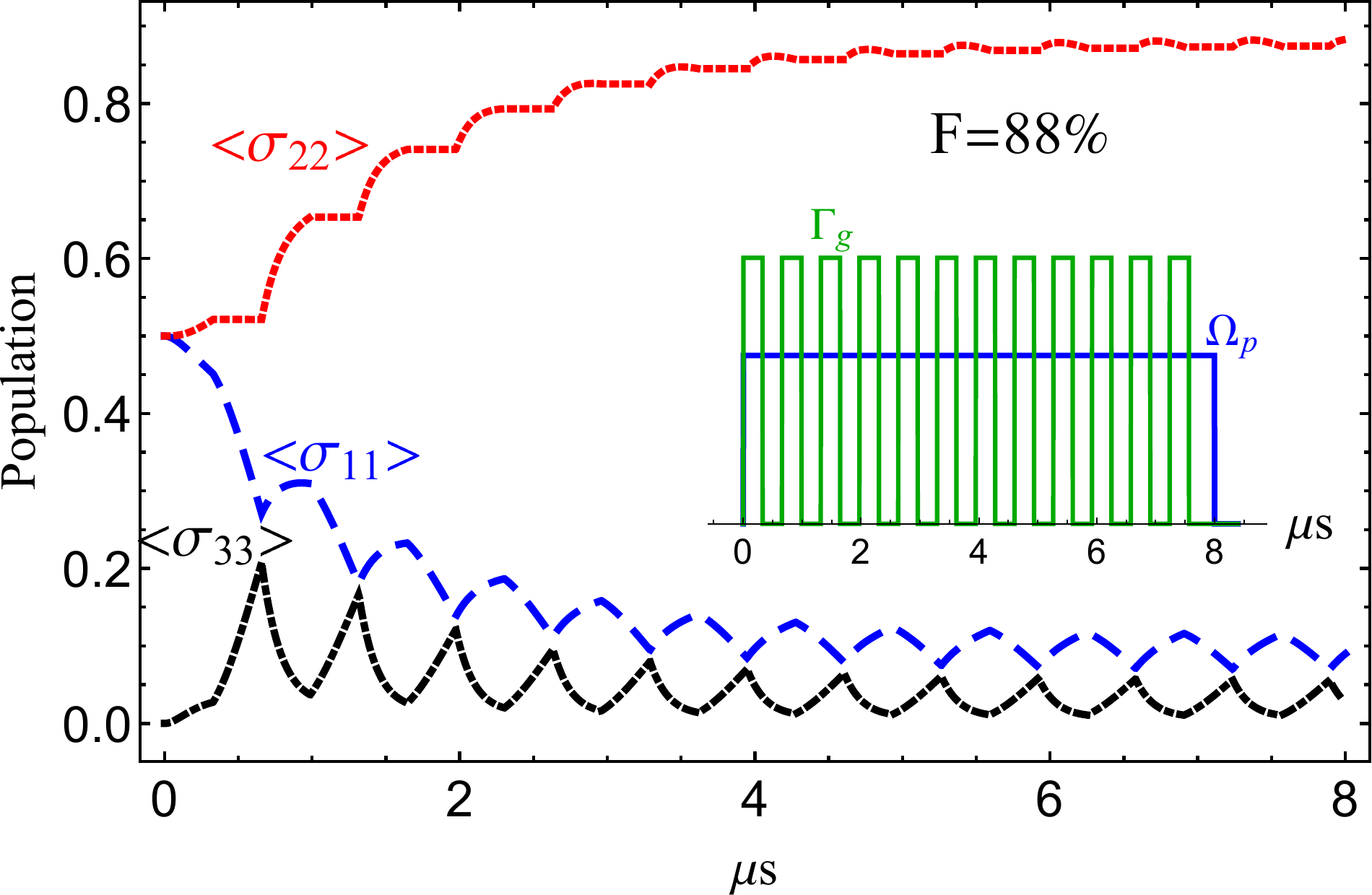}
\caption{ Successful preparation of the nuclear spin $\ket{\downarrow}$ starting from a thermal state, in the presence of the green laser $\Gamma_g=5$ MHz and $\Delta=0$. Inset: pulse sequence with the pump field ($\Omega_p/2\pi=0.5$ MHz) and the green laser. }
\label{fig2_polarization}
\end{figure}

We observed that the longitudinal relaxation considerably harms the preparation. For this reason, the laser is pulsed with an on-time of approximately $300$ ns (see Inset of Fig. \ref{fig2_polarization}). In this way the effect of the depolarization is minimised. We note that this process is similar to the one presented in Ref \cite{Jamonneau}. After 8 $\mu$s it is possible to prepare the nuclear spin state $\ket{m_n= \downarrow}$ ($\ket{\psi_2}$) with a fidelity of $88\%$. To prepare the other spin state $\ket{m_n= \uparrow}$ ($\ket{\psi_1}$), the pump field must be replaced by the Stokes field $\Omega_S(t)$. Therefore, the nuclear spin can be polarized on either $\ket{\psi_1}$ or $\ket{\psi_2}$ state. It is worthwhile noticing that when larger Rabi frequencies can be reached, for instance $\Omega_0/2\pi > 3$ MHz, a square laser (similar to $\Omega_p$) rather than a pulsed laser, leads to similar fidelities but in shorter times.  

The evolution was estimated using the Master equation

\begin{equation}\label{ME}
\frac{d\rho}{dt} = -i[\tilde{H},\rho] +  \mathcal{L}_k^{\uparrow}(\rho) + \mathcal{L}_k^{\downarrow}(\rho) + \mathcal{L}_1^{D} + \mathcal{L}_3^{D} + \mathcal{L}_{12}^{dep},
\end{equation}
where the Lindblad operators are given by
\begin{eqnarray}
\mathcal{L}_k^{\downarrow} &=& \sum_{k=1,2} \frac{\Gamma_{3k}}{2}(1+n_{th})(2\sigma_{k3}\rho\sigma_{3k}-\sigma_{33}\rho -\rho\sigma_{33}) , \nonumber\\
 \mathcal{L}_k^{\uparrow} &=&\sum_{k=1,2} ( \frac{\Gamma_{3k}}{2}(n_{th})(2\sigma_{3k}\rho\sigma_{k3}-\sigma_{kk}\rho -\rho\sigma_{kk}), \nonumber \\
  \mathcal{L}_{l}^{D} &=& \frac{\Gamma_{l D}}{2}(2\sigma_{ll}\rho\sigma_{ll}-\sigma_{ll}\rho -\rho\sigma_{ll}), \nonumber\\
   \mathcal{L}_{12}^{dep} &=& \sum_{i\neq j}\frac{\Gamma_{n,opt}}{2}(2\sigma_{ij}\rho\sigma_{ji}-\sigma_{ii}\rho -\rho\sigma_{ii}),\nonumber
\end{eqnarray}
and $\Gamma_{31}$ ($\Gamma_{32}$) is the decay rate from state $\ket{3}$ to state $\ket{\psi_1}$ ($\ket{\psi_2}$) due to the effect of thermal phonons and it is intrinsic to the system so we will consider it throughout this paper. For practical considerations we set the upward and downward transitions to be of the same order, contrary to what is assumed in the optical domain where the upward transitions are usually neglected. These terms account for the $T_1$ relaxation process that thermalise the electronic spin of the NV centre, although modelling this process is more complex\cite{Jarmola}. The relaxation time $T_1$ is of the order of few ms \cite{Jarmola}. The decoherence mechanisms bellow are optically induced by the green laser and will be only considered in particular cases. The fourth and fifth terms in Eq.(\ref{ME}) correspond to pure dephasing (energy conserving) process over the states $\ket{\psi_1}$ and $\ket{\psi_3}$. In our system, the strength of the dephasing term $\Gamma_{1D}$ is given by two components, the optically induced transverse relaxation $\Gamma_{D,opt}$ and the intrinsic decoherence time $T_{2n}^{\ast}$ of the Carbon-$13$, which is of the order of tens or hundreds of $m$s \cite{Gurudev}. On the other hand, $\Gamma_{3D}$ is related to the decoherence time $T_2^\ast$ of the electronic spin, which is of the order of a few $\mu$s \cite{Childress,Maze,Taminiau}. The last term corresponds to a depolarization channel over the ground state, i.e. $\lbrace i,j \rbrace =1,2$\cite{Jamonneau}.

Other sources of decoherence coming from the presence of substitutional Nitrogen, known as $P_1$ centres, are sample dependent and the interaction between $P_1$s and NVs depends on the external magnetic field \cite{Hanson}. These sources are not considered here but they might be important at values of the external magnetic field for which both species are on resonance \cite{Jarmola}.

We now analyse the robustness of the present approach. In Fig. \ref{fig3_polarization}a we consider the effect of different Rabi frequencies $\Omega_0$ and the single photon detuning $\Delta$. We note that by increasing $\Omega_0$ the fidelity increases. However, by increasing the single photon detuning, the preparation is completely destroyed and cannot be overcome with the green laser.
We also analyse the effect of both longitudinal relaxation and laser-induced decay rates in Fig. \ref{fig3_polarization}b. The former ($\Gamma_{n,opt}$), rapidly hurts the fidelity, reducing the applicability of our approach. For $\Gamma_{n,opt}=0$ and $\Gamma_{g}=5$ MHz the fidelity reaches $99.4\%$. The latter ($\Gamma_{g}$) has a plateau, such that for  $\Gamma_{g}\geq 3$ MHz the fidelity is above $99\%$ (at $\Gamma_{n,opt}=0$).

\begin{figure}[h]
\centering 
\includegraphics[scale=0.3]{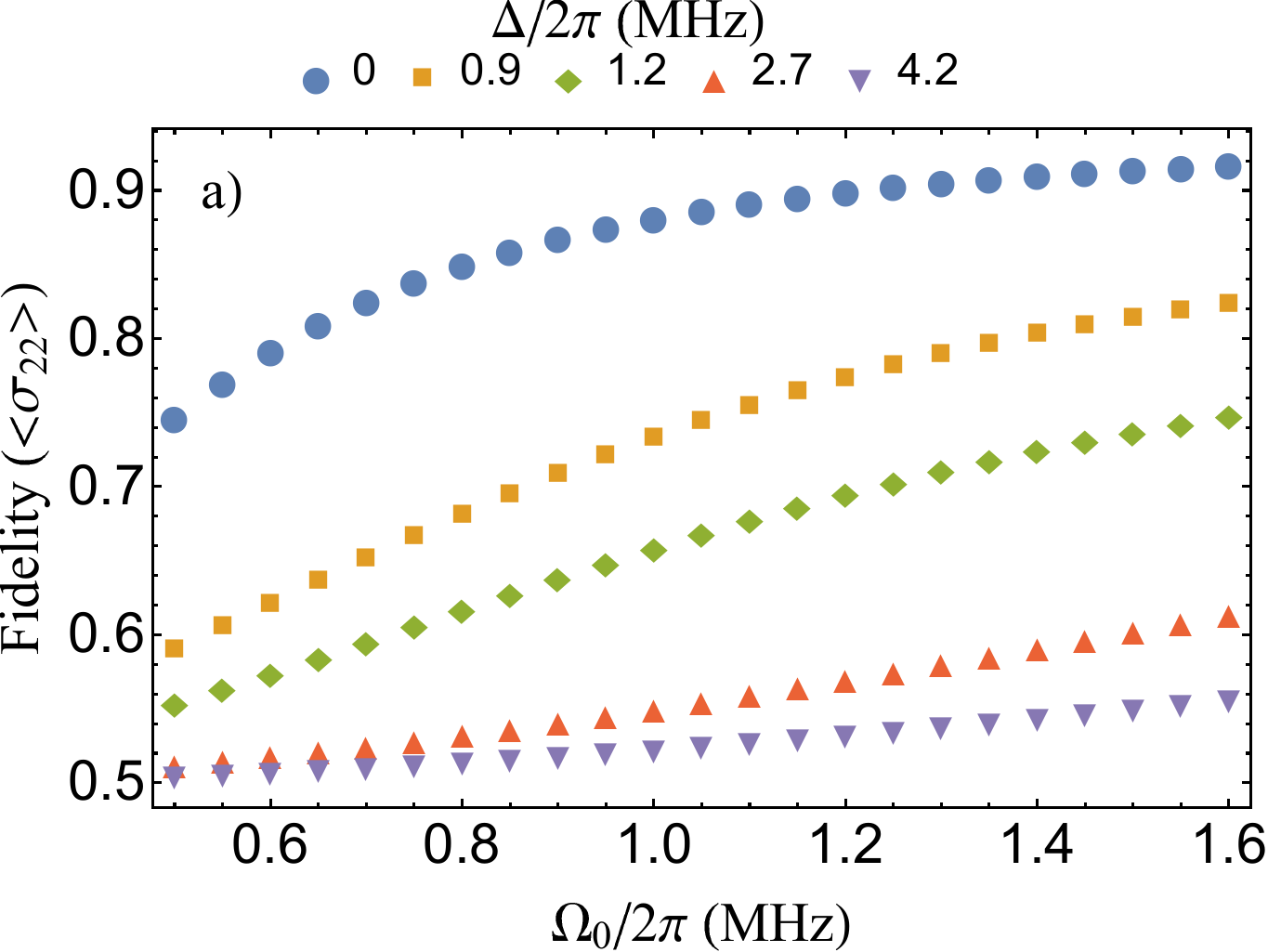}
\includegraphics[scale=0.3]{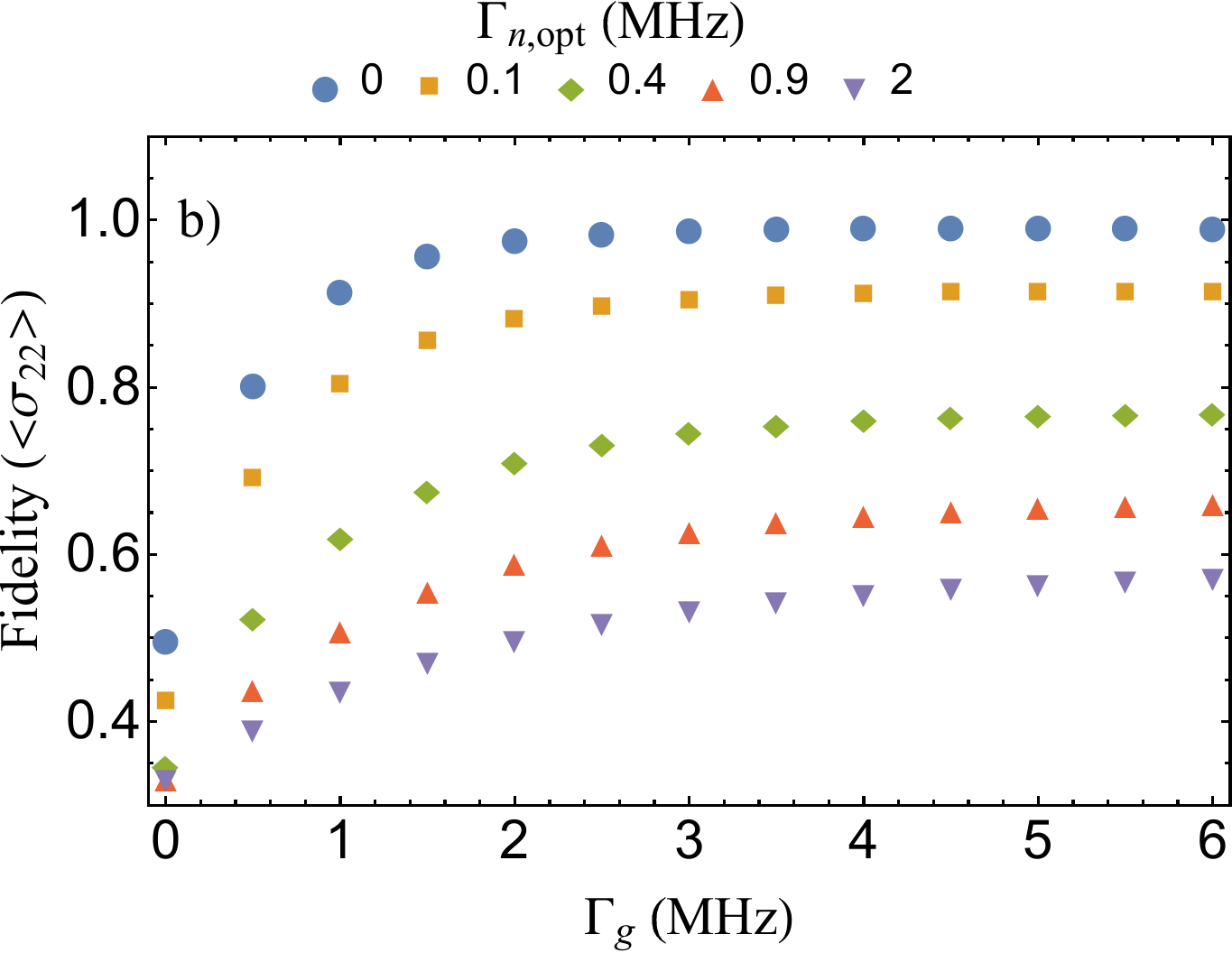}
\caption{Fidelity of the preparation. a) As a function of the single photon detuning $\Delta$ and Rabi frequency $\Omega_0$. $\Gamma_{g}=5$ MHz, $\Gamma_{n,opt}=150$ kHz and $\Gamma_{1D}=1$ MHz. b) As a function of longitudinal $\Gamma_{n,opt}$ and laser-induced $\Gamma_{g}$ decay rates of the nuclear spin. $\Omega_0/2\pi=1$ MHz, $\Delta =0$ and $\Gamma_{1D}=1$ MHz. }
\label{fig3_polarization}
\end{figure}

In general, in $\Lambda$-schemes, the robustness in terms of pure dephasing on the ground state $\Gamma_{1D}$ is detrimented. However, this can be overcome in the presence of the decay $\Gamma_g$ induced by the green laser. We observed that for a large transverse relaxation rate $\Gamma_{1D}=1$ MHz the population in the radiative state $\ket{\psi_3}$ grows, but the fidelity holds. It is not a surprise that the same conclusion can be extended to the original STIRAP process, where one starts from an already polarized state ($\ket{\psi_1}$). As known, the STIRAP is fragile to ground state decoherence ($\Gamma_{1D}\neq 0$). For instance, the population transfer decreases to $\mean{\sigma_{22}}=0.4$ with a pure dephasing noise of strength $\Gamma_{1D}=1$ MHz, same parameters as in Fig. \ref{fig4_STIRAP}. Nevertheless, the presence of the green laser enhances the population transfer, reaching $96\%$ of success for $\Gamma_{n,opt}=0$ and $82\%$ for $\Gamma_{n,opt}=150$ kHz. Similar schemes have been previously studied to prevent the effect of pure dephasing on the ground state for STIRAP \cite{Scala,Mathisen,Wang}.

\subsection{Fast manipulation of the nuclear spin via STIRAP}

A complete toolbox for controlling a $^{13}C$ nuclear spin requires the ability to prepare a coherent superposition state, which commonly implies the use of a radio frequency (RF) field \cite{Jiang:2009,Brown2011}. Under a limited amplitude of the RF field, such approaches require long times compared to the time required to manipulate electronic spins. For example, the time needed for preparing a nuclear superposition at a given Rabi frequency scales as $T_{\pi/2}^n=(\gamma_e/\gamma_n )T_{\pi/2}^e$ where $T_{\pi/2}^e = \pi/(2\Omega_0)$. For  $T_{\pi/2}^e = 1/4$ $\mu$s, $T_{\pi/2}^n$ gives $654$ $\mu$s. 
On the contrary, in the lambda system presented here the nuclear spin rotate between states $\ket{\psi_1}$ and $\ket{\psi_2}$ following the faster electronic transitions by taking advantage of the non-nuclear-spin-preserving transitions ($\ket{\psi_2}\leftrightarrow \ket{\psi_3}$ and $\ket{\psi_1}\leftrightarrow \ket{\psi_3}$).

\begin{figure}[ht]
\centering 
\includegraphics[scale=0.4]{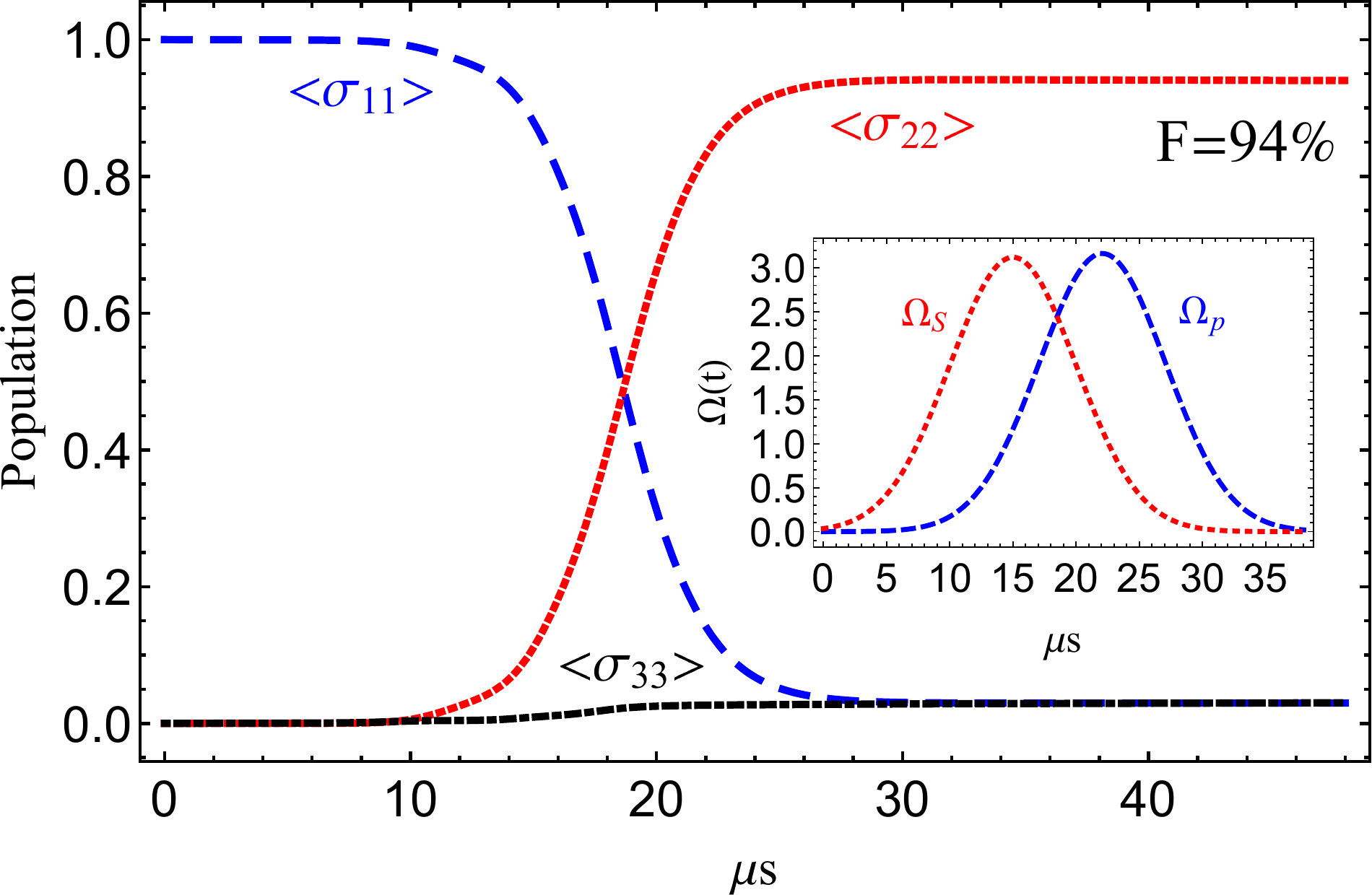}
\caption{Population transfer through Stimulating Raman Adiabatic Passage (STIRAP) of the nuclear spin $\ket{m_n=\uparrow}$ ($\mean{\sigma_{11}}$) to $\ket{\downarrow}$ ($\mean{\sigma_{22}}$) with $94\%$ of success. $\sigma=5$ $\mu$s, $\Gamma_{1D}=0$, $\Gamma_{3D}=1$ MHz, $\Gamma_{31}\approx \Gamma_{32}\approx 9\times10^{-5}$ MHz, $\Delta=0$. Inset: STIRAP pulse sequence, Stokes pulse ($\Omega_S$) precedes the pump pulse ($\Omega_p$). }
\label{fig4_STIRAP}
\end{figure}

Consider, for example, on a $\Lambda$-scheme with initial population on state $\ket{\psi_1}$, a STIRAP process for which the Stokes pulse precedes the pump pulse (see inset of Fig. \ref{fig4_STIRAP}). Such pulse order is known as conterintuitive pulse sequence \cite{Klein}. At the initial time ($t\rightarrow -\infty$), $\theta = 0$, while at the end of the interaction ($t\rightarrow +\infty$), $\theta = \pi/2$. Fig. \ref{fig4_STIRAP} shows how the population evolves from an initially prepared state $\ket{\psi_1}$ to state $\ket{\psi_2}$.  The transfer time is of the order of $30$ $\mu$s with a fidelity of $94\%$ , which is limited by the lack of adiabaticity and the decoherence in the excited state $\Gamma_{3D}$. This is considerably shorter than the time required for an RF field which directly couples to the nuclear spin for the same Rabi frequency, which is about $1.3$ ms. This time can be decreased by increasing the RF power at expenses of heating the sample. The fidelity can be improved by following a more adiabatic evolution. For example, by increasing the width of the pulses so that $\sigma=11$ $\mu$s, the fidelity reaches $97\%$ for a transfer time of $65$ $\mu$s.

\begin{figure}[ht]
\centering 
\includegraphics[scale=0.4]{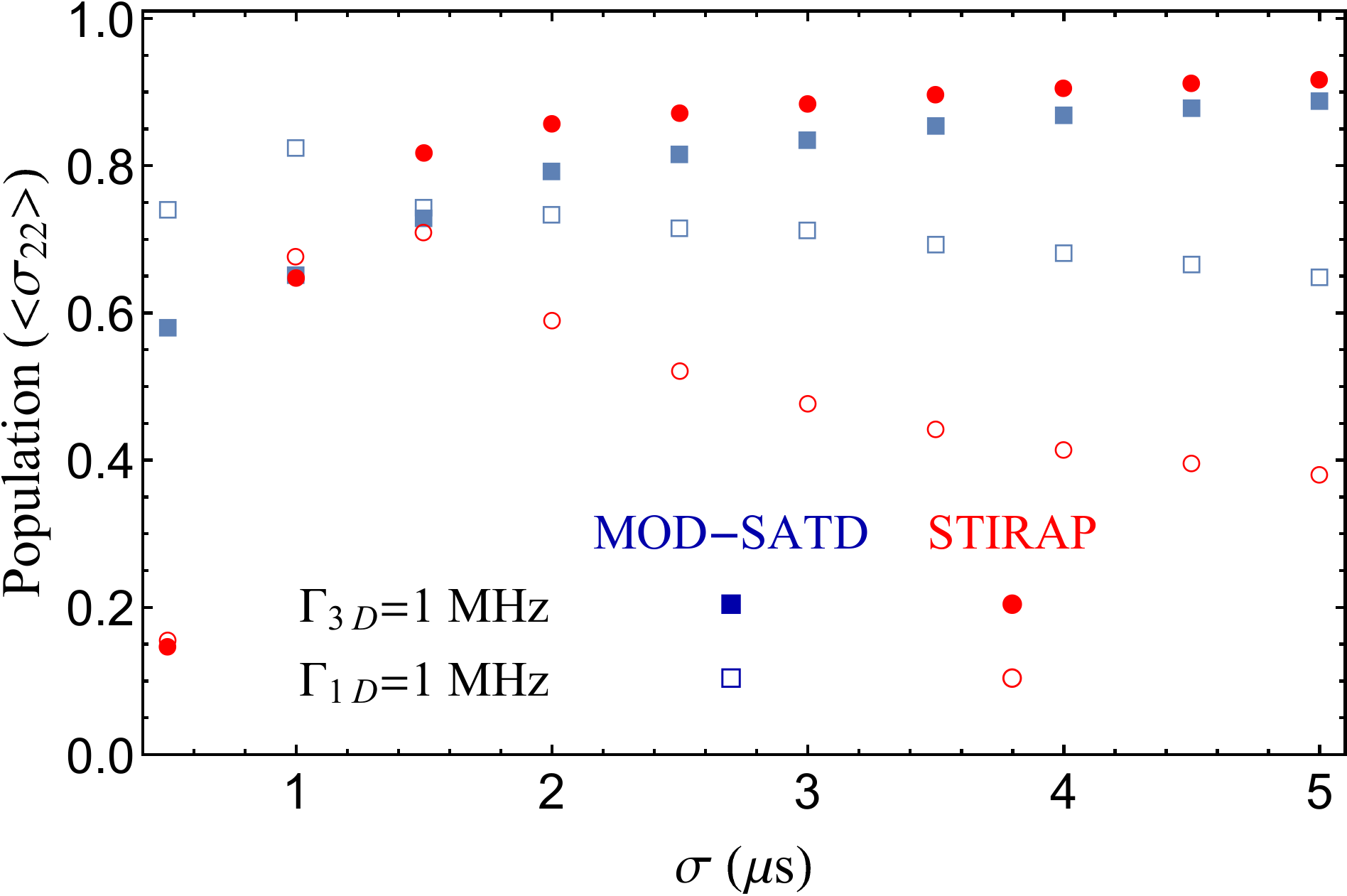}
\caption{MOD-SATD protocol speed up STIRAP, outperforming the latter in the case where only dephasing losses in the ground state are present, $\Gamma_{1D}=1$ MHz and $\Gamma_{3D}=0$. When losses appear only in the excited state, $\Gamma_{1D}=0$ and $\Gamma_{3D}=1$ MHz, the protocol fails in being superior to STIRAP. For STIRAP we set $t_d=\sqrt{2}\sigma$, while for MOD-SATD $t_d$ has been optimized for each $\sigma$.  }
\label{fig5_SATD}
\end{figure}

Hence, the manipulation of a nuclear state can be performed an order (or even two orders) of magnitude faster than conventional methods. In the same way, a superposition state can be created by applying a fraction of the STIRAP sequence\cite{Timoney,Webster}. For example, it takes $19$ $\mu$s to create the state $1/\sqrt{2}(\ket{\uparrow} - \ket{\downarrow})$ with fidelity $96\%$. Even more, these times might be further improved by hastening the STIRAP process. For this purpose several protocols exist \cite{Baksic,Zhou,Masuda,He}, where the main idea consists of bypassing the adiabatic condition, counteracting the effect of the loss of adiabaticity with an external control (auxiliary field). Toward this goal, we focus on the work proposed recently by Baksic \textit{et al.} \cite{Baksic}, termed MOD-SATD (modified superadiabatic transitionless driving). The aspects of this approach have been detailed in Appendix \ref{Appendix:MOD}. We observe that in the absence of dephasing noise, MOD-SATD outperforms STIRAP, allowing to reach higher population transfer in shorter times (not shown here). However, in the presence of dephasing noises in the excited ($\Gamma_{3D}$) and ground ($\Gamma_{1D}$) states, a trade off between these two approaches shows up, separating their range of effectiveness. In Fig. \ref{fig5_SATD} we calculate the population that reaches the target state ($\mean{\sigma_{22}}$) as a function of the pulse width $\sigma$, which controls the effective duration of the protocol. Notice that in the presence of $\Gamma_{3D}=1$ MHz and for not too short dynamics ($\sigma>1~\mu$s) STIRAP prevails as a good protocol, because the MOD-SATD deliberately occupies the excited state, suffering of strong decoherence. For a short dynamics ($\sigma<1 ~\mu$s) the performance interchanges and STIRAP deteriorates considerably. Nevertheless when the dephasing noise is only present in the ground state ($\Gamma_{1D}=1$ MHz), MOD-SATD leads the population transfer.

\subsection{Ramsey spectroscopy and geometric phase}

\begin{figure}[ht]
\centering 
\includegraphics[scale=0.18]{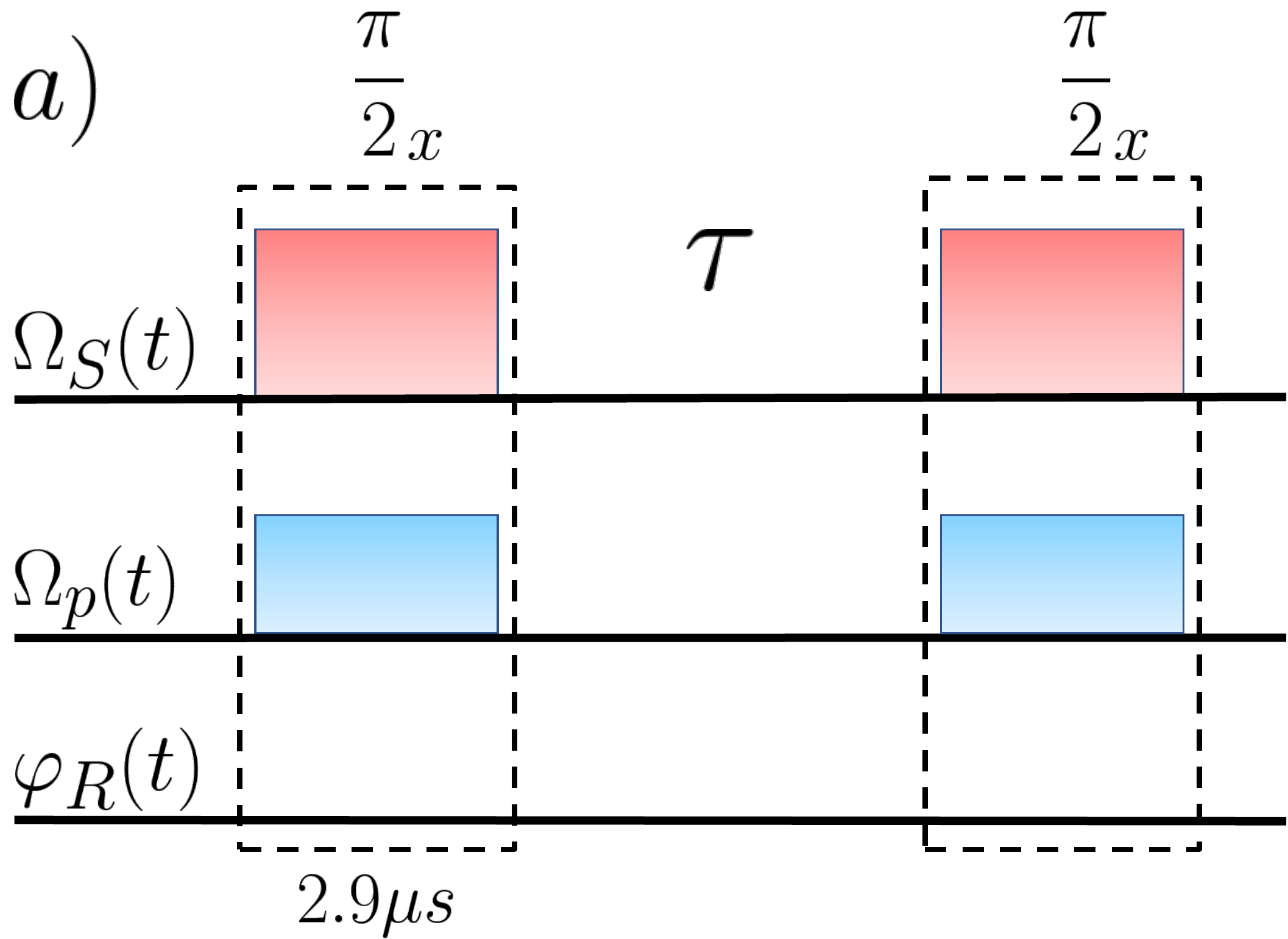}
\includegraphics[scale=0.22]{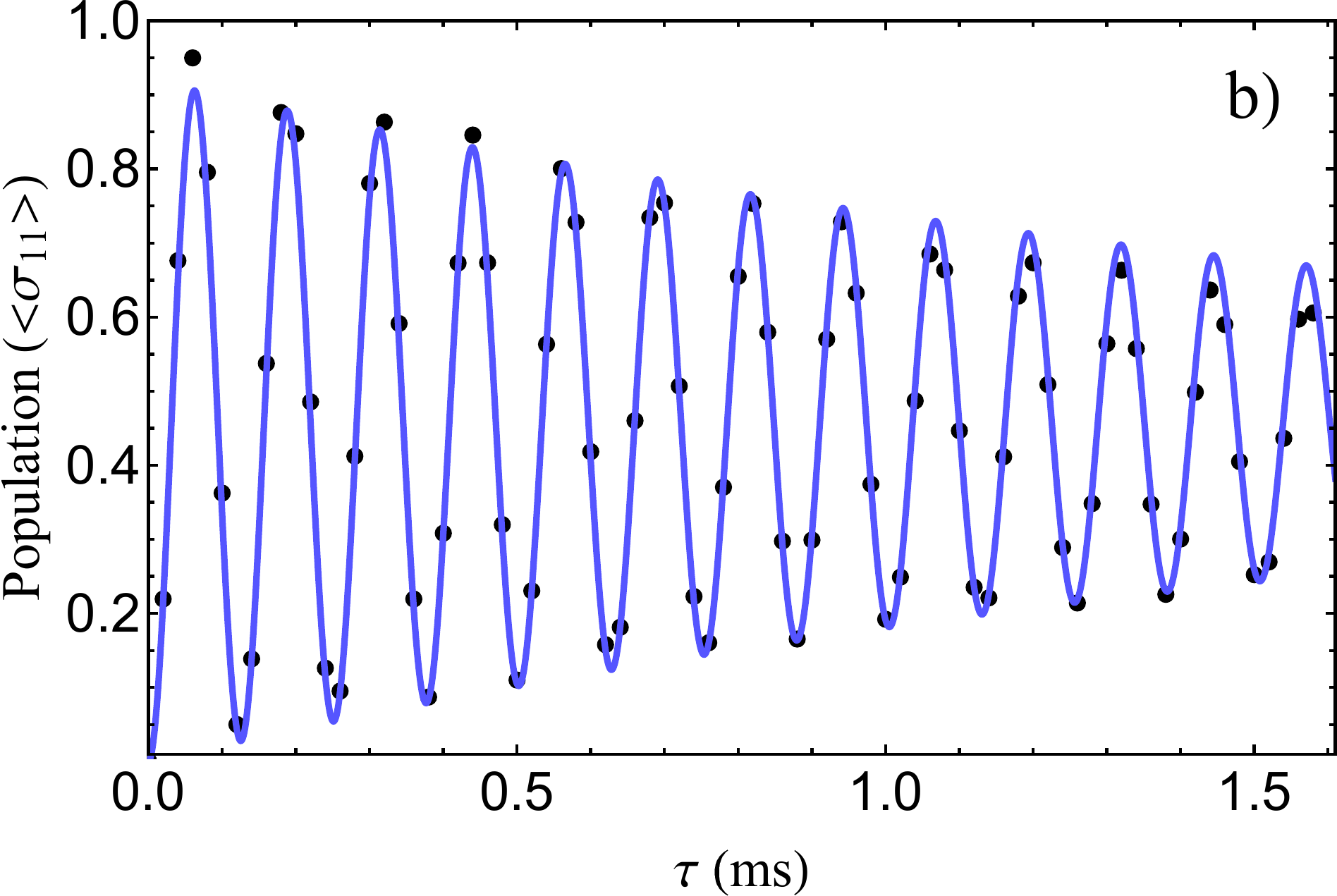}
\includegraphics[scale=0.15]{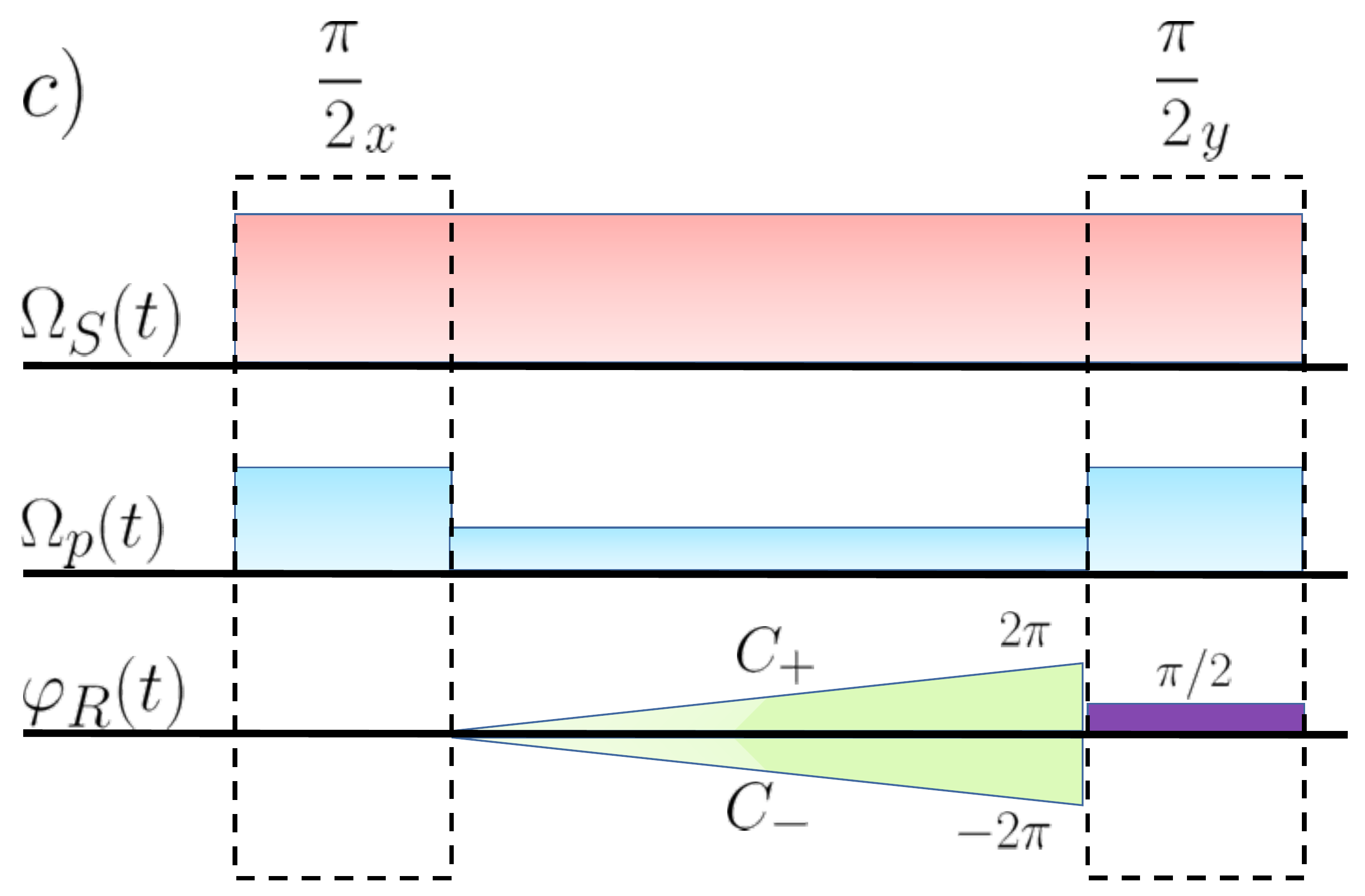}
\includegraphics[scale=0.22]{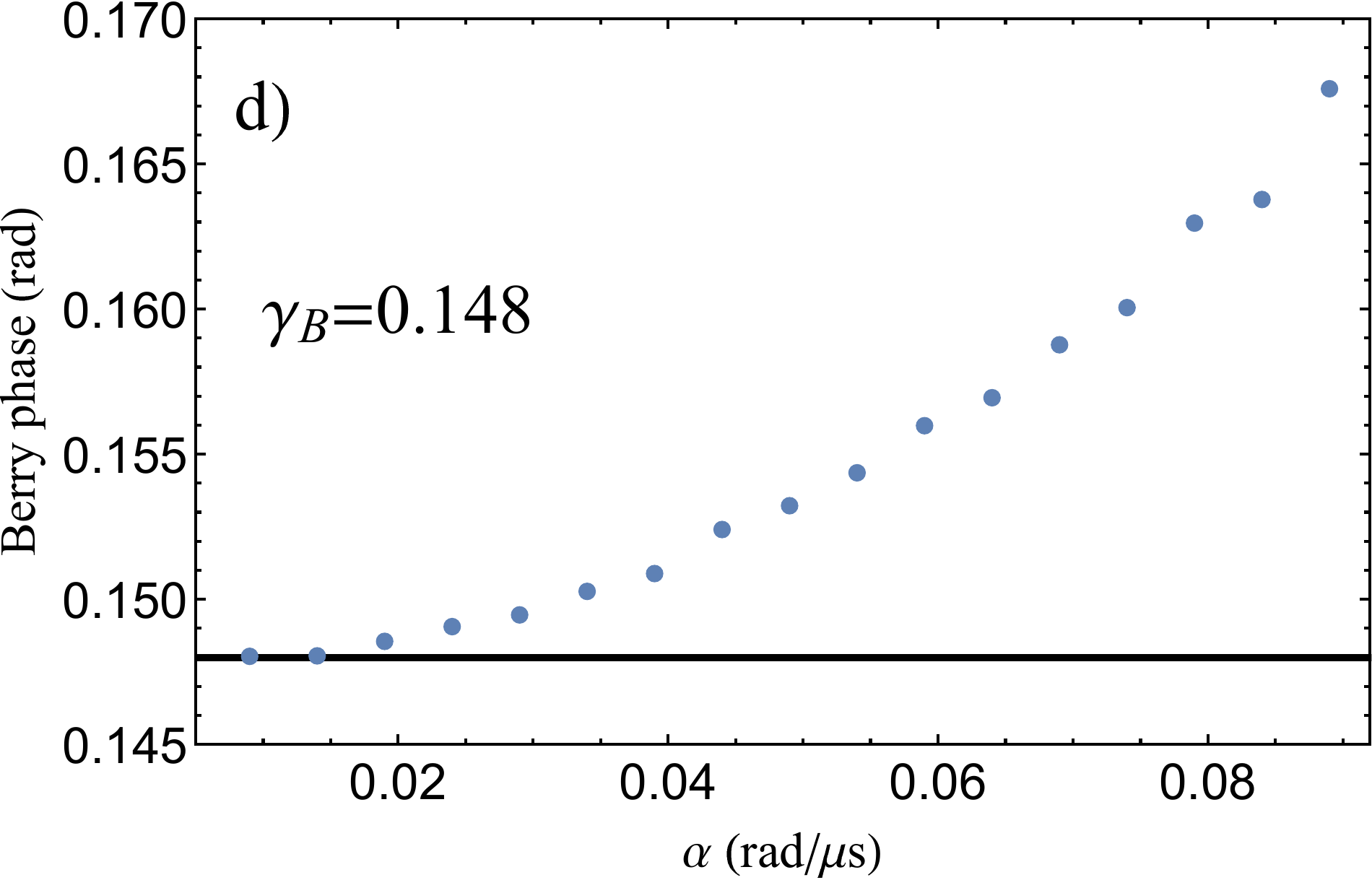}
\caption{(color online) a) Pulse sequence used for Ramsey spectroscopy, $\Omega_S/2\pi=1.76$ MHz and $\Omega_p/2\pi=1.61$ MHz. b) Example of Ramsey interferometry using the nuclear spin of the Carbon-$13$. Analytical result (solid) and simulation (dots) agrees. $\Gamma_g=0$, $\Gamma_{1D}=1$ kHz, $\Delta/2\pi= 16$ MHz and $\delta/2\pi=0.05/2\pi$ MHz. The solid curve is a fit to the numerical data using a function $a_0(1 -e^{-(\Gamma_{1D}/2) \, \tau} \cos (\delta \tau))$. The constant $a_0\approx 0.46$ differs from the ideal value $1/2$ because of the losses during the preparation of the superposition and further application of the pulses to recover the initial state. c) Pulse sequence for measuring the Berry's phase. d) Variation of the Berry's phase as a function of the ramp speed for varying the relative phase between the Stoke and pump pulses. As the ramp speed increases, the calculated value for the phase deviates from the expected value (horizontal line) due to the loss of adiabaticity.}
\label{fig6_Ramsey}
\end{figure}

In this section we show how to implement Ramsey spectroscopy for metrology purposes and for measuring the geometric phase acquired by a nuclear spin nearby to a NV centre. To gain further insight on the dynamics of the ground state, we adiabatically eliminate the excited state $\ket{\psi_3}$ and arrive to the following effective Hamiltonian

\begin{eqnarray}
H &=&\left(-\frac{\vert\Omega_p\vert^2}{4\Delta}+\frac{\vert\Omega_S\vert^2}{4\Delta}-\delta\right)\frac{\sigma_z}{2} -\mathcal{R}\left\{\frac{\Omega_p^\ast\Omega_S}{4\Delta}\right\}\sigma_x \nonumber \\&-& \mathcal{I}\left\{\frac{\Omega_p^\ast\Omega_S}{4\Delta}\right\}\sigma_y \label{H_effective},
\end{eqnarray}
where $\sigma_z=\sigma_{11}-\sigma_{22}$, $\sigma_x=\sigma_{12}+\sigma_{21}$ and $\sigma_y=-i\sigma_{12}+i\sigma_{21}$. Note that the shape of the pulses $\Omega_S$ and $\Omega_p$, and their relative phase can be arranged to arbitrarily move the nuclear state on the Bloch sphere. Without loss of generality, we consider that one of the Rabi frequencies has a time-dependent phase $\Omega_p = \vert\Omega_{p} \vert e^{i\varphi_R(t)} $ with $\vert\Omega_{p} \vert = \Omega_{13} \cos(\vartheta/2)^2$. The $\sigma_z$ component can be controlled by replacing the Gaussian profiles of the microwave pulses by two overlapped rectangular pulses of different amplitude, e.g. $\Omega_S>\Omega_p$. Thus, by taking $\varphi_R=0$ ($\varphi_R=\pi/2$), the spin rotates only around x (y). Let us consider first a $(\pi/2)x$ pulse over the initial state $\ket{\psi_1}$, as depicted in Fig. \ref{fig6_Ramsey}a. This pulse prepares the state $1/\sqrt{2}(\ket{\psi_1}+i\ket{\psi_2})$ in approximately $3$ $\mu$s with a fidelity over $98\%$. This superposition freely evolves for a time $\tau$, subject to pure dephasing losses given by $\Gamma_{1D}$, until we apply another $(\pi/2)x$ pulse in order to map the phase differences acquired during the free evolution to population differences. We notice that the precession during the free evolution comes from the two photon detuning, which has been set to $\delta/2\pi=0.05/2\pi$ MHz for illustration purposes. The resulting signal is plotted in Fig. \ref{fig6_Ramsey}b, as a function of the precession time $\tau$. The slow decay is a consequence of the decoherence of the nuclear spin at a rate of $\Gamma_{1D}$. These results are particularly useful for sensing low frequency components of external magnetic fields due to the low decoherence rate of nuclear spins.

Finally, we explore the different phases acquired by the nuclear spin. The effective Hamiltonian (\ref{H_effective}) can be rewritten as, $H_R = \vec{\Omega}\cdot\vec{\sigma}$, where $\vec{\sigma}=\lbrace \sigma_x,\sigma_y ,\sigma_z \rbrace$ and $\vec{\Omega}=\lbrace -\Omega_R\cos(\varphi_R) ,- \Omega_R\sin(\varphi_R), \Omega_z \rbrace$, with $\Omega_R = \vert\Omega_p\vert \vert\Omega_S\vert / 4\Delta$ and $\Omega_z=(\vert\Omega_S\vert^2 - \vert\Omega_p\vert^2 - 4\Delta\delta)/8\Delta $. Notice that this Hamiltonian is suitable for measuring Berry's phase \cite{Berry}, provided that the phase $\varphi_R(t)$ is adiabatically varied such that $\vec{\Omega}$ completes a closed path. For $\varphi_R$ varied from $0$ to $2\pi$, the acquired geometric phase is $\pm 2\pi(1-\cos(\theta_R))/2$, where the sign $\pm$ refers to the opposite phases acquires by the eigenstates and $\cos(\theta_R)=\Omega_z/\sqrt{\Omega_R^2 + \Omega_z^2}$. Therefore, $\gamma_B = 2\pi(1-\cos(\theta_R))$ is the relative geometric phase that equals the solid angle enclosed by the cone that traces $\vec{\Omega}(t)$ around the z axis.

One can use different approaches to directly observe this phase, for instance, STIRAP pulse sequence \cite{Yale:2016,Molmer} and spin-echo pulse sequence \cite{Leek}. The latter leads directly to the relative geometric phase by canceling the dynamical phase, while the former takes advantages of the evolution of the dark state with corresponding zero energy. However, we will obtain the Berry phase through the Ramsey scheme mentioned above. First, we prepare a superposition as illustrated in Fig. \ref{fig6_Ramsey}c. Then, we start varying the phase $\varphi_R(t)$ adiabatically. This adiabaticity requires that $\dot{\varphi}_R\sin(\theta_R)/(2\vert \vec{\Omega}\vert) \ll 1$ \cite{Leek}, for which we set $\varphi_R= \pm \alpha t$, with $\alpha=0.006$ and the sign $\pm$ refers to the direction of the path $C_\pm$. A closed path is obtained for an evolution time $T=2\pi/\alpha$. We leaved the amplitude of the Stoke pulse invariant during the whole process, $\Omega_S/2\pi=1.76$ MHz, while the pump pulse is reduced to $\Omega_p/2\pi=0.16$ MHz during the adiabatic evolution.

By traversing the path in one direction ($C_+$), the eigenstates of the Hamiltonian acquire a total (dynamical plus geometric) relative phase $\phi_+=\delta_d(t)+\gamma_B$, while in the opposite direction ($C_-$) the phase is $\phi_-=\delta_d(t)-\gamma_B$, where $\delta_d(t)$ stands for the dynamical phase. Note that $\delta_d(t)$ is independent of the direction of the path. Hence, repeating the process for each path, allows us to obtain the geometric (Berry) phase as $\gamma_B=(\phi_+ - \phi_-)/2 $, where $\phi=\arctan[(\cos(\chi)\mean{\sigma_x}-\sin(\chi)\mean{\sigma_z})/\mean{\sigma_y}]$ and $\chi=\arctan[\Omega_R/ \Omega_z]$. The expectation value of $\mean{\sigma_y}$ can be found by applying the final $\pi/2$ rotation around the x axis (see Fig. \ref{fig6_Ramsey}c) and measuring the population in the excited state, $P_e=1/2( 1 + \mean{\sigma_y})$. A similar procedure reveals $\mean{\sigma_x}$. Finally, $\mean{\sigma_z}$ can be directly obtained as $P_e=1/2( 1 - \mean{\sigma_z})$ without applying any final rotation. Fig. \ref{fig6_Ramsey}d shows the calculated Berry phase and the adiabatic expected value as a function of the ramp speed $\alpha$. It can be seen how the calculated Berry phase deviates from the expected value as the ramp speed is increased due to the loss of adiabaticity.

\section{Conclusions}

Based on the theory of Stimulated Raman Adiabatic Passage (STIRAP), we proposed a feasible scheme to speed up the manipulation of a nearby Carbon-13 by exploiting the anisotropy of the hyperfine interaction, which enables us to implement a $\Lambda$-configuration that allows nuclear spin to follow the faster electronic transitions. We found that the time needed for preparing a superposition or make a spin-flip can be considerably smaller than the time used in conventional methods involving RF fields. Moreover, we showed that the modification of STIRAP known as MOD-SATD increases the fidelity for short pulses in the presence of losses. A protocol for preparing a nuclear spin state from an initially thermal state is also discussed in detail. This can be used for achieving total control of a nuclear spin state. As an example, we show how to perform Ramsey spectroscopy for either metrology purposes or to measure the different, dynamical and geometric, phases acquired by the nuclear spin nearby an NV centre in diamond. 

\vspace{10pt}

\textbf{Acknowledgments} \newline
We thank M. Orszag and B. Seifert for fruitful discussion. RC acknowledges the financial support from Fondecyt Postdoctorado No. 3160154. JRM acknowledges support from Conicyt-Fondecyt grant  No. {1141185}. GH would like to ackowledge funding by the French National Research Agency (ANR) through the project SMEQUI.

\appendix

\section{Dependence of the $\Lambda$-Configuration with $B_z$}\label{Appendix_B}   

One of the most striking feature of this $\Lambda$-system, is that the whole configuration depends on the magnetic field $B_z$. To see this, one can notice that the transition frequencies $T_{31}=E_3-E_1$ and $T_{32}$ are functions of $B_z$ \cite{Dreau}. Then, the difference in energy of these two transition is $T_{31}-T_{32}=\gamma_n B_z$. For instance, for a low magnetic field ($B_z\approx 20$ G), this difference can be neglected, leading to $T_{31}=T_{32}$. The transition frequencies are not the only elements depending on $B_z$. The intensity of these two transitions are also $B_z$-dependent. The relative intensity between forbidden (not nuclear spin conserving transition $T_{32}$) and allowed transitions is given by $\tan^2(\vartheta/2)$ \cite{Dreau}, with $\vartheta$ defined in Eq.(\ref{vartheta}). When $\vert A_{zz}-\gamma_n B_z \vert$ and $A_{ani}$ have the same order of magnitude, all the transitions can be observed. If $A_{zz}=\gamma_n B_z$, i.e., when $\vartheta=\pi/2$, the amplitudes of the transitions $T_{31}$ and $T_{32}$ are identical. This case happens at $B_z\approx 950$ G. We took $A_{ani}=0.51$ MHz and $A_{zz}=1.02$ MHz, from the experiments \cite{Dreau}. The last case is $\vert A_{zz}-\gamma_n B_z \vert \gg A_{ani}$, where only nuclear spin conserving transitions can be observed. In order to consider this effect, the Rabi frequency $\Omega_p$ ($\Omega_S$) as well as the decay rate $\Gamma_{31}$($\Gamma_{32}$) will be weighted by $\cos(\vartheta/2)^2$ ($\sin(\vartheta/2)^2$). 

\section{MOD-SATD protocol}\label{Appendix:MOD}

The MOD-SATD protocol is a generalization of the counterdiabatic approach, that eliminates the flaw of connecting the initial and target state by introducing modifications only to the original Stokes and pump fields \cite{Baksic,Zhou}. These corrections in the fields naturally appear  when transforming the original adiabatic basis to a dressed state basis, that reproduces the STIRAP outcome but without the constraint of an adiabatic evolution. Following Ref. \cite{Baksic}, we parametrize the pump and Stokes field as

\begin{equation}
\Omega_p=-\cos^2(\vartheta/2)\Omega(t)\sin\theta, \hspace*{0.2cm} \Omega_S=\sin^2(\vartheta/2)\Omega(t)\cos\theta,
\end{equation}

and corrects the angles and amplitude such that 

\begin{eqnarray}
\theta &\rightarrow & \theta(t)-\arctan[\frac{g_x(t)}{\Omega(t)+g_z(t)}], \\
\Omega(t)  &\rightarrow & \sqrt{(\Omega(t)+g_z(t))^2+g_x^2(t)},
\end{eqnarray}

where for our Gaussian pulses we set $g_x(t)=\dot{\mu}$, $g_z(t)=-\Omega-\dot{\theta}/\tan(\mu)$ and
\begin{eqnarray}
\theta(t)&=&\arctan[\exp(2t_dt/\sigma^2)], \\
\Omega(t)&=& \Omega_0\exp(-\frac{t^2+t_d^2/4}{\sigma^2})\sqrt{2\cosh(t_dt/\sigma^2)},\\
\mu(t) &=& -\arctan[\frac{\dot{\theta}}{g(t)/\sigma +\Omega(t)}].
\end{eqnarray}
 
$g(t)=A/\cosh(\zeta t)$ with $A=1/40$ and $\zeta=9/10\sigma$ was selected according to Ref. \cite{Baksic}.

\end{document}